\documentclass[12pt,preprint]{aastex}
\newcommand{\pks}{PKS~0637--752}
\newcommand{\axaf}{\mbox{\em Chandra\/}}

\slugcomment{\today}
\begin{document}

\title{\axaf\  Observations of Magnetic Fields and Relativistic
Beaming in  Four Quasar Jets}
\author{D.A. Schwartz,\altaffilmark{1} 
H.L. Marshall,\altaffilmark{2} J.E.J. Lovell,\altaffilmark{3} D.W.
Murphy,\altaffilmark{4} G.V. Bicknell,\altaffilmark{5} 
M.~Birkinshaw,\altaffilmark{1,6}
J.~Gelbord,\altaffilmark{2}  M. Georganopoulos,\altaffilmark{7,8}
L. Godfrey,\altaffilmark{3,5}
D.L. Jauncey,\altaffilmark{3}  
E.S.~Perlman\altaffilmark{7}
D. M. Worrall\altaffilmark{1,6} }
\altaffiltext{1}{Harvard-Smithsonian Center for Astrophysics, 60
  Garden Street, Cambridge, MA 02138} 
\altaffiltext{2}{Kavli Institute for Astrophysics and Space Research,
  Massachusetts Institute of Technology, 77 Massachusetts Avenue,
  Cambridge, MA 02139} 
\altaffiltext{3}{CSIRO Australia Telescope National Facility, PO Box
  76, Epping NSW 1710, Australia} 
\altaffiltext{4}{Jet Propulsion Laboratory, 4800 Oak Grove Drive,
  Pasadena, CA 91109} 
\altaffiltext{5}{Research School of Astronomy and Astrophysics,
  Australian National University, Cotter Road, Weston Creek, Canberra,
ACT72611}
\altaffiltext{6}{Department of Physics, University of Bristol, Tyndall
Avenue, Bristol BS8 1TL, UK} 
\altaffiltext{7}{Department of Physics, Joint Center for Astrophysics,
  University of Maryland-Baltimore County, 1000 Hilltop Circle,
  Baltimore, MD 21250}
\altaffiltext{8}{NASA's Goddard Space Flight Center, Mail Code 660,
  Greenbelt, MD 20771}
\email{das@head-cfa.harvard.edu}

\begin{abstract}

We discuss the physical properties of four quasar jets imaged with the
\emph{Chandra} X-ray Observatory in the course of a survey for X-ray
emission from radio jets \citep{Marshall05}.  These objects have
sufficient counts to study their spatially resolved properties, even
in the 5 ks survey observations. We have acquired Australia Telescope
Compact Array data with resolution matching \emph{Chandra}. We have
searched for optical emission with Magellan, with sub-arcsecond
resolution.  The radio to X-ray spectral energy distribution for most
of the individual regions indicates against synchrotron radiation from
a single-component electron spectrum.  We therefore explore the
consequences of assuming that the X-ray emission is the result of
inverse Compton scattering on the cosmic microwave background.  If
particles and magnetic fields are near minimum energy density in the
jet rest frames, then the emitting regions must be relativistically
beamed, even at distances of order 500 kpc from the quasar.  We
estimate the magnetic field strengths, relativistic Doppler factors,
and kinetic energy flux as a function of distance from the quasar core
for two or three distinct regions along each jet. We develop, for the
first time, estimates in the uncertainties in these parameters,
recognizing that they are dominated by our assumptions in applying the
standard synchrotron minimum energy conditions.  The kinetic power is
comparable with, or exceeds, the quasar radiative luminosity, implying
that the jets are a significant factor in the energetics of the
accretion process powering the central black hole.  The measured
radiative efficiencies of the jets are of order 10$^{-4}$.

\end{abstract}


\section{INTRODUCTION}

Following the remarkable discovery of an X-ray luminous, 100 kpc scale
jet in \pks\ \citep{Schwartz00, Chartas00}, we have embarked on a
survey to investigate the occurrence and properties of such
systems. Initial goals were to assess the frequency of detectable
X-ray fluxes from radio-bright jets, to locate good targets for
detailed imaging and spectral 
followup studies, and where possible to test models of the X-ray
emission by measuring the broad-band, spatially resolved,  spectral
energy distributions 
(SED) of jets from the radio through the optical to the
X-ray band \citep{Marshall05, Schwartz03a,Schwartz03b}. With an X-ray
detection of 12 jets out of the first set of 20 observed,
\citep{Marshall05}, the survey has been successful in
meeting these objectives. For four of these jets we have 40 to 130 total
X-ray counts in our 5 ks observations.  This suffices to construct
broad-band spectral energy distributions, from which we can estimate
magnetic fields, particle densities, Doppler beaming factors and
kinetic fluxes for independent, spatially distinct emitting regions,
using the models of synchrotron radiation and inverse Compton (IC)
scattering on the cosmic microwave background (CMB), which we adopt
below \citep{Schwartz03a,Schwartz03b}.  Deeper \axaf\  as well as HST
observations have been approved for all these sources
(\citet{Perlman04}, and in preparation). 

In parallel work, \citet{Sambruna02, Sambruna04} have undertaken a survey of
17 jetted radio quasars with Chandra and HST, with 10 exhibiting
at least one knot in the Chandra images. We will compare our results
with theirs in Section~\ref{param}.

Our survey is described in \citet{Marshall05} (hereafter Paper I). We
selected objects from two parent samples for which radio maps had been
obtained with $\sim$ 1\arcsec\ -- 2\arcsec\ resolution at 1-10 GHz: a
VLA sample (Dec $\ge$ 0\arcdeg) of flat spectrum quasars with core
flux densities $S_{5 \rm {GHz}} \geq 1$ Jy \citep{Murphy93} and an
ATCA survey of flat spectrum Parkes quasars (Dec $<$ -20\arcdeg)
\citep{Lovell97}, with core flux densities $S_{2.7 \rm {GHz}} \geq
0.34$ Jy.  Selection was then made for objects with radio jets which
extend beyond 2\arcsec\ from the core.  The survey is comprised of
sources for which we either anticipated extended structures with
significant X-ray fluxes, based on scaling the extended 5 GHz flux
using the X-ray to radio flux ratio of \pks, (subsample ``A''), or
else selected by the morphological criteria of a one-sided linear jet
(subsample ``B''). PKS~0920-397 and PKS~1202-262 meet the ``A''
criterion, and PKS~0208-512, PKS~1030-357, and PKS~1202-262 meet the
``B'' criterion (Paper I).  The definition of our parent samples by
flat radio core spectra tends to select powerful sources with
one-sided relativistic pc scale jets beamed toward our line of
sight. The 5 or 2.7 GHz selection frequency emphasizes the jet rather than
lobe emission.

Section~\ref{obs} presents the X-ray data and defines distinct spatial
regions for further analysis. Section~\ref{spectral} gives the broad
band spectral energy distribution, and Section~\ref{param} the
physical properties deduced by assuming that IC/CMB produces the X-ray
emission. Section~\ref{sec:discussion} discusses implications of the
IC/CMB mechanism, and mentions alternate emission mechanisms. In
Appendix~\ref{app:errors} we estimate the systematic uncertainties in
the magnetic fields and Doppler factors which we derive, and in
Appendix~\ref{app:kinetic} we present the basis for our calculation of
the flux of kinetic energy carried by the jets.

\section{OBSERVATIONS OF THE JETS}
\label{obs}

\begin{figure}[t]
\plottwo{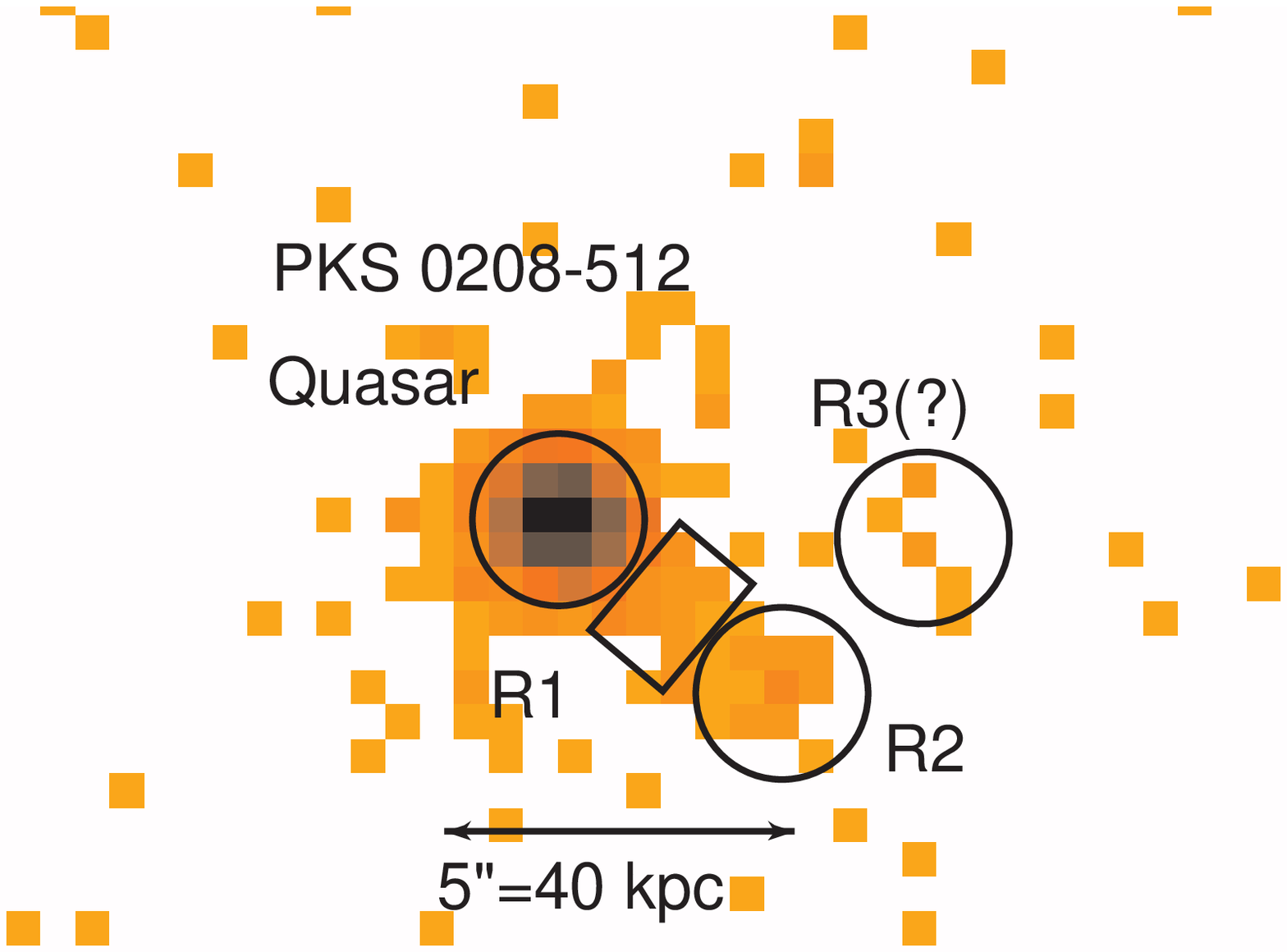}{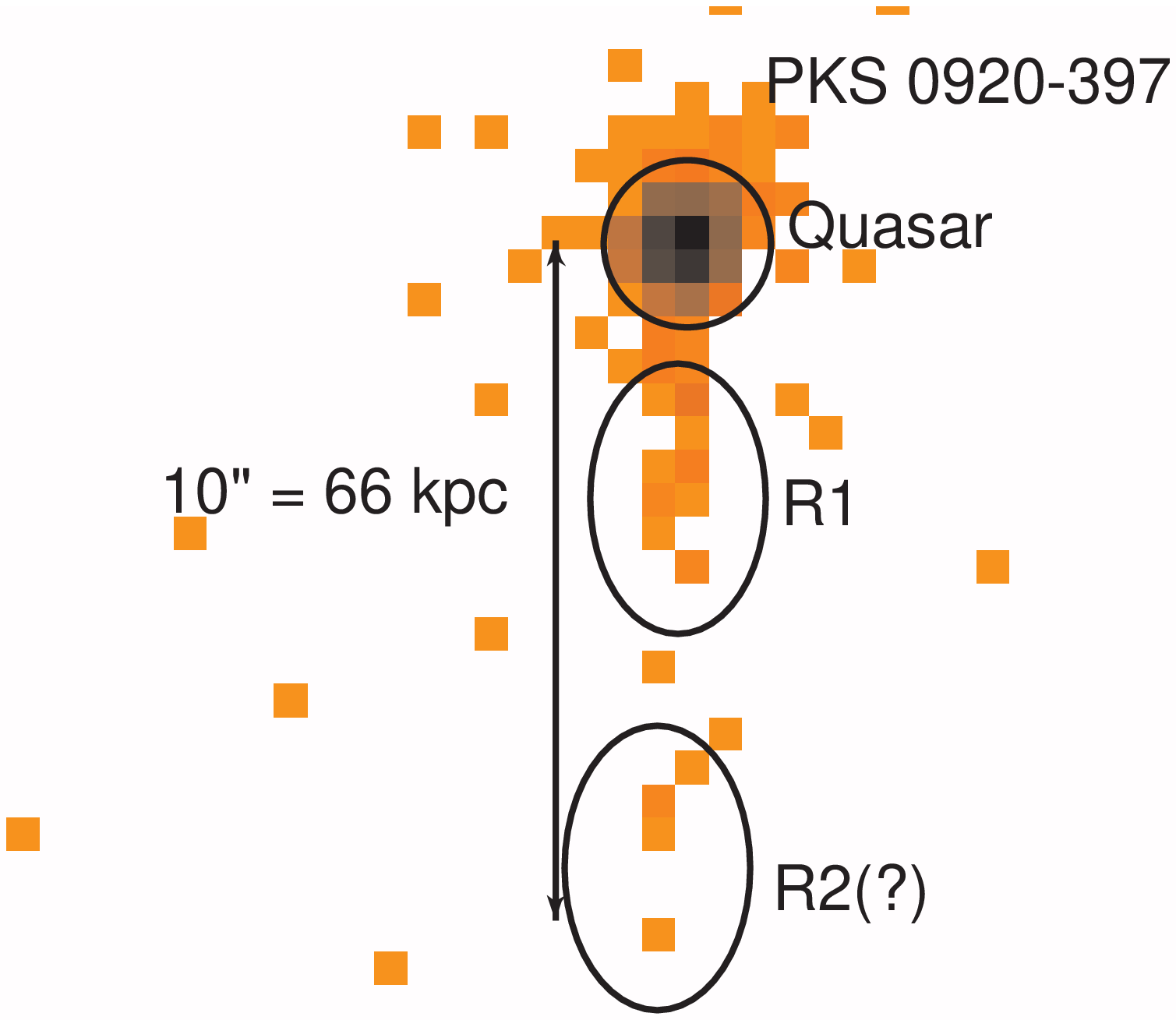}
\plottwo{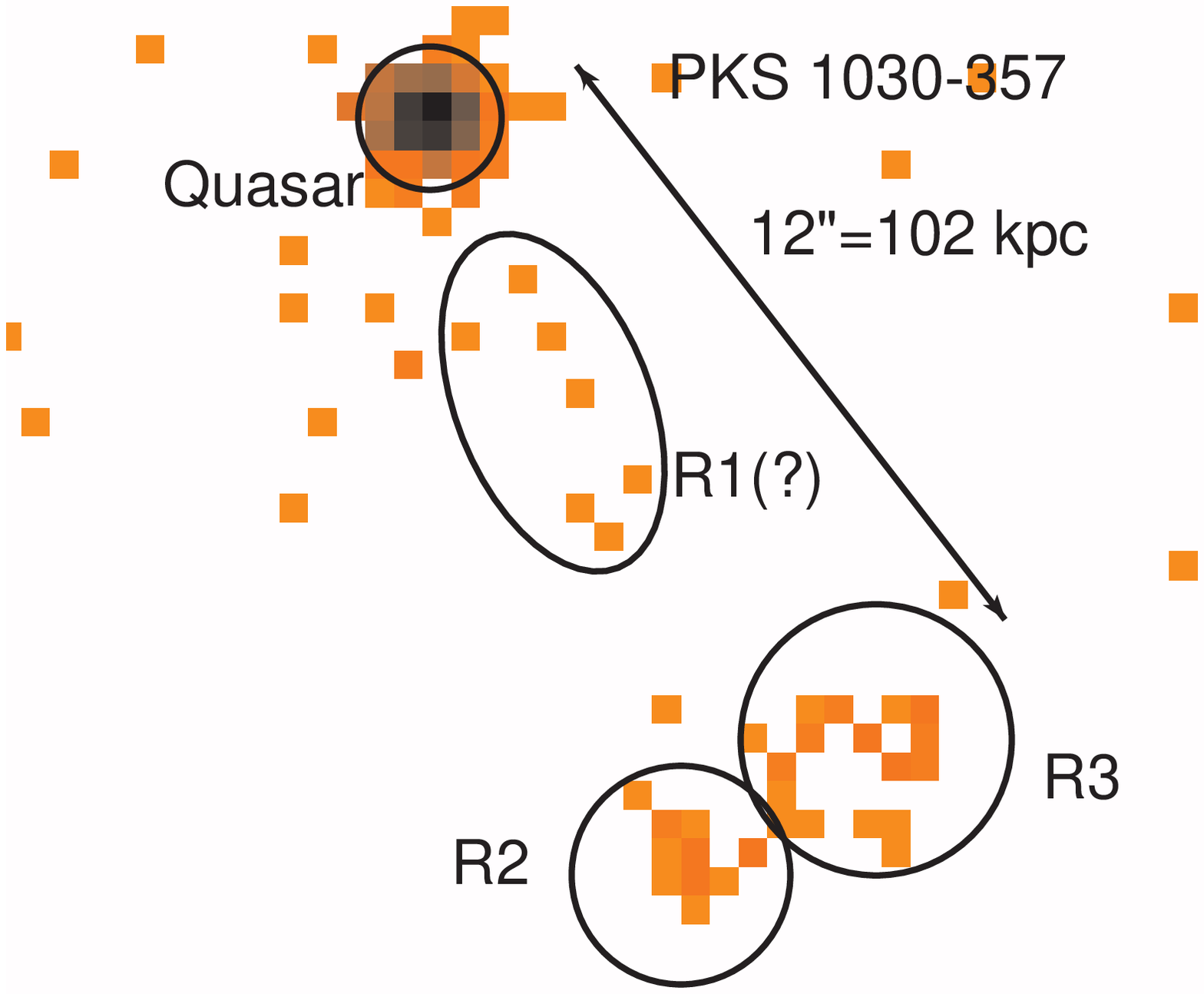}{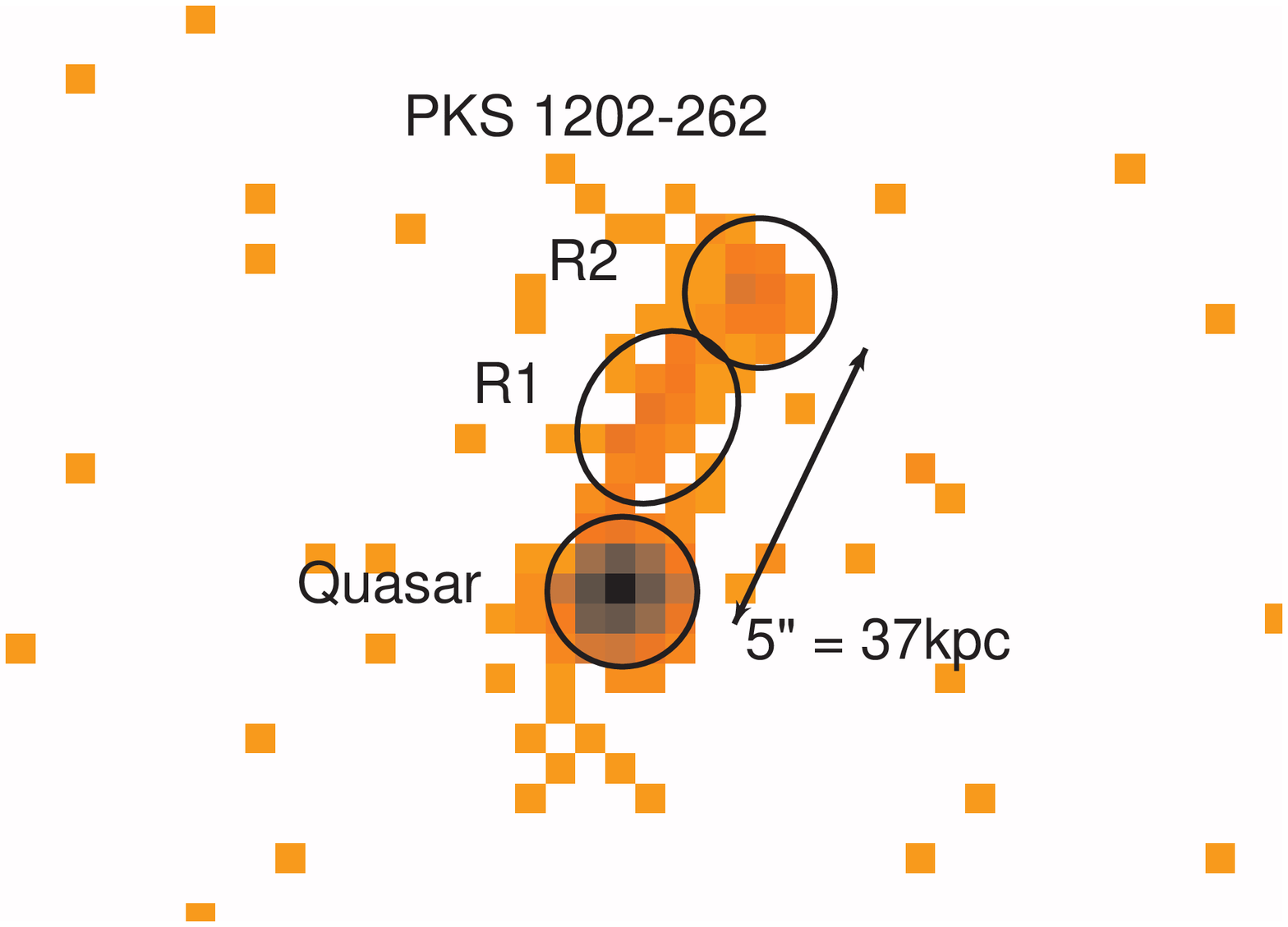}
\epsscale{0.8}
 \caption{\label{fig:regions} ACIS-S X-ray counts, 0.5 to 7 keV,
  binned into 0\farcs492 pixels. The regions marked \emph{R1, R2, R3}
  are used for the SED plots and the analysis of spatially resolved
  parameters. The circles centered on the quasars have a 1\farcs23
  radius, and enclose approximately 95\% of the flux. The faintest
  color is one count per pixel. See text for definition of the marked
  regions. The straight lines give projected distances in the plane of
  the sky, at the redshift of each object. All images are in celestial
  coordinates with N up and E left, and cover 21\arcsec
  $\times$~15\arcsec.}
\end{figure}

We observed the four quasars listed in Table~\ref{tab:data} for about
5 ks, using ACIS-S.  Data for the jets as a whole and X-ray results
for the quasar cores appear in \citet{Marshall05}.  We generally used
the 1/4 sub-array mode to minimize pileup of the quasar core, but used
the 1/8 subarray for PKS 0208-512 due to its greater flux.  We
requested roll angles such that the radio jet projected at least
30\arcdeg\ away from the readout streak.  Paper I (figure 1a, 1g, 1h,
1k), shows the overlay of the 8.64 ATCA GHz radio contours on the
X-ray images. Figure~\ref{fig:regions} shows the X-ray images of the
four jets, and defines the regions used for the joint X-ray/radio
spatial analysis.  The jet regions were manually placed on the X-ray
images, and therefore involve a certain amount of subjectivity. The
regions are labeled \emph{R} with numbers increasing away from the
quasar. They are intended primarily to be regions larger than the
instrumental resolution, and distinct from the quasar core, so we
could derive independent model parameters characteristic of distinct
volumes within each jet.
\clearpage
\clearpage
The definitions of the spatial regions included a minimum of 5 X-ray
photons. Background counts are expected to be less than 0.1 in any
region, and are ignored.  Uncertainties are dominated by the X-ray
statistics and model assumptions, as we shall discuss in
Appendix~\ref{app:errors}, 
therefore we do not believe the exact region definitions affect any of
the present conclusions.  However, we do note with a question mark in
Table~\ref{tab:data} 
three regions for which there are less than 10 X-ray counts. The
reality of detection and spatial location of these, and therefore
their association with the radio jet, must be regarded as less
certain.

Table~\ref{tab:data} gives the observed X-ray and radio data for each
region. To derive the 1-keV X-ray flux density, $f_{1\, keV}$, and
2--10 keV rest frame luminosity, L$_{x}$, we assume a power-law energy
index of $\alpha$=0.7, where $f_{\nu} \propto \nu^{-\alpha}$. Our
model that the radio and X-rays arise from the same simple power-law
population of electrons implies that this is also the index of the
radio emission.\footnote{Note that the GHz emitting electrons have
Lorentz factors $\approx$ 30 times larger than the electrons
responsible for the 
IC X-rays} For 3 of the quasars in Table~\ref{tab:data} this is
consistent with the spectral index from 4.8 to 8.6 GHz,
$\alpha_{4.8}^{8.6}$, as tabulated in column 9,\footnote{Except for
PKS~1202-262, the 4.8 GHz data are at poorer resolution than the 8.6
GHz images, and this produces additional errors in
$\alpha_{4.8}^{8.6}$} within errors whose effect will be discussed in
Appendix~\ref{app:errors}. For PKS 0208-512 the indices deviate
markedly from $\alpha$=0.7; nevertheless we perform the formal
calculation using the same index for comparison with the other
objects. For this source, we have contamination of the R1
4.8 GHz flux density by the core, and the R2 and R3
regions may be a hotspot and extended lobe, and so require different
modeling.  Our deeper \emph{Chandra} and 20 GHz ATCA observations of
this source will address these issues in the future.

The X-ray properties reported here differ slightly from those in Paper
I, because that paper considered the complete jet region, while in
this work we omit some counts outside the regions marked in
Figure~\ref{fig:regions}.  The angular sizes of the radio and X-ray
regions are calculated by subtracting in quadrature the 1\farcs2 FWHM
of the 8.6 GHz images given in Paper I from the dimensions of each
region.  The regions are all resolved in the 8.6 GHz beam.  We
consider the regions as cylinders of angular length, $\theta_{l}$, and
diameter $\theta_{d}$.  We use a flat, accelerating cosmology, with
H$_0$=71 km s$^{-1}$ Mpc$^{-1}$, $\Omega_{\rm m} = 0.27$, and
$\Omega_{\Lambda}=0.73$.

\clearpage

\begin{deluxetable}{rcrrrrrrrrr}    
\footnotesize
\tablewidth{0pt}
\tablecaption{{Observations of Quasars, and X-Ray Jet Regions\label{tab:data}} }
\tablecolumns{11}
\tablehead{
\colhead{{PKS Name\tablenotemark{a}}} &
\colhead{{}} &
\colhead{Exposure} &
\colhead{{X-ray}} &
\colhead{f$_{\rm{1\, keV}}$} &
\colhead{{}} &
\colhead{{f$_{\rm{4.8\, GHz}}$}}&
 \colhead{{f$_{\rm{8.6\, GHz}}$}} & 
\colhead{{$\alpha_{4.8}^{8.6}$\tablenotemark{c}}} &
\colhead{{$\theta_l$\tablenotemark{d}}} & 
\colhead{{$\theta_d$\tablenotemark{d}}} \\

\colhead{{(region)}} &
\colhead{{z\tablenotemark{a}}} &
\colhead{Time [ks]} &
\colhead{Counts} &
\colhead{[nJy]} &
\colhead{{L$_{\mathrm{x}}$\tablenotemark{b}}} &

\colhead{{[mJy]}} &
\colhead{{[mJy]}}&
\colhead{{}} &
 \colhead{{[arcsec]}} & 
\colhead{{[arcsec]}}
}

\startdata
{
0208-512\tablenotemark{e}  }&{0.999 }&{5.014 }&{1556 }&{319
}&{8.9 }&{3270}&{3020  }&0.1&{\nodata}&{\nodata}\\
(R1) & & &16  &3.3 &0.091 &{\nodata}  &12.8 &{\nodata}  & $<$0.50&2.14 \\
(R2) & & &23  &4.7 &0.131 &25.3  &25.5&-0.0  &1.93 &1.01 \\
(R3)? & & & 7  &1.4 &0.040 &16.6  & 7.7&1.3 &1.05 &1.99 \\
 & & & & & & & & & \\

{0920-397\tablenotemark{e}}&{0.591 }&{4.466 }&{520 }&{ 120 }&{0.973}&{
 1740 }&{1570  }&0.2&{\nodata}&{\nodata}\\
(R1) & & &19 &4.4 &0.036 &\nodata &38.4&{\nodata} &3.68 &0.62 \\
(R2)? & & & 5&  1.2&0.009 & 105 &66.8&0.8 &1.56 &1.30 \\
 & & & & & & &  & & \\

{1030-357\tablenotemark{e} }&{1.455 }&{5.029}&{ 395 }&{ 80.8 }&{ 5.39 }&{ 241 }&{ 173
}&0.6&{\nodata}&{\nodata} \\
(R1)? & & &7 & 1.4& 0.095& 28&22.4&0.4 &5.88 &0.47 \\
(R2) & & &17 & 3.5& 0.232&22.2 &14.5&0.7  &1.25 &0.42 \\
(R3) & & &28 & 5.7&0.382 &53.8 & 36.2&0.7 &2.89 &0.84 \\
 & & & & & & &  & & \\

{1202-262\tablenotemark{e}}&{0.789}&{5.074} &{ 754 } &{ 153 } &{2.44 }
&{464}&{482}&-0.1&{\nodata}&{\nodata}\\ 
(R1) & & &57 &11.6 &0.185 &32.1 &22.7&0.6 &2.91 &0.35 \\
(R2) & & &50 & 10.1&0.162 &45.7 & 31.2&0.7 &0.91 &0.75 \\
\enddata
\tablecomments{Region designations followed with a question mark
  contain fewer than 10 X-ray counts, so their physical association
  with the radio emission is   not  certain.}
\tablenotetext{a}{{From the NED database, operated by JPL for NASA}}
\tablenotetext{b}{{Rest frame 2 --10 keV luminosity, in units of
    10$^{45}$ergs s$^{-1}$, assuming isotropic radiation}} 
\tablenotetext{c}{{Spectral index from 4.8 GHz to 8.6 GHz, defined as
    S$_{\nu} \propto \nu^{-\alpha}$}}
\tablenotetext{d}{{Angular length, $\theta_{l}$, and diameter,
    $\theta_{d}$, of the regions in Figure~\ref{fig:regions}, after
    subtracting the radio beam 1\farcs2 FWHM in quadrature}}
\tablenotetext{e}{{Quasar core properties}}
\end{deluxetable}

\clearpage
\section{SPECTRAL DISTRIBUTIONS}
\label{spectral}

\begin{figure}[h]
\plottwo{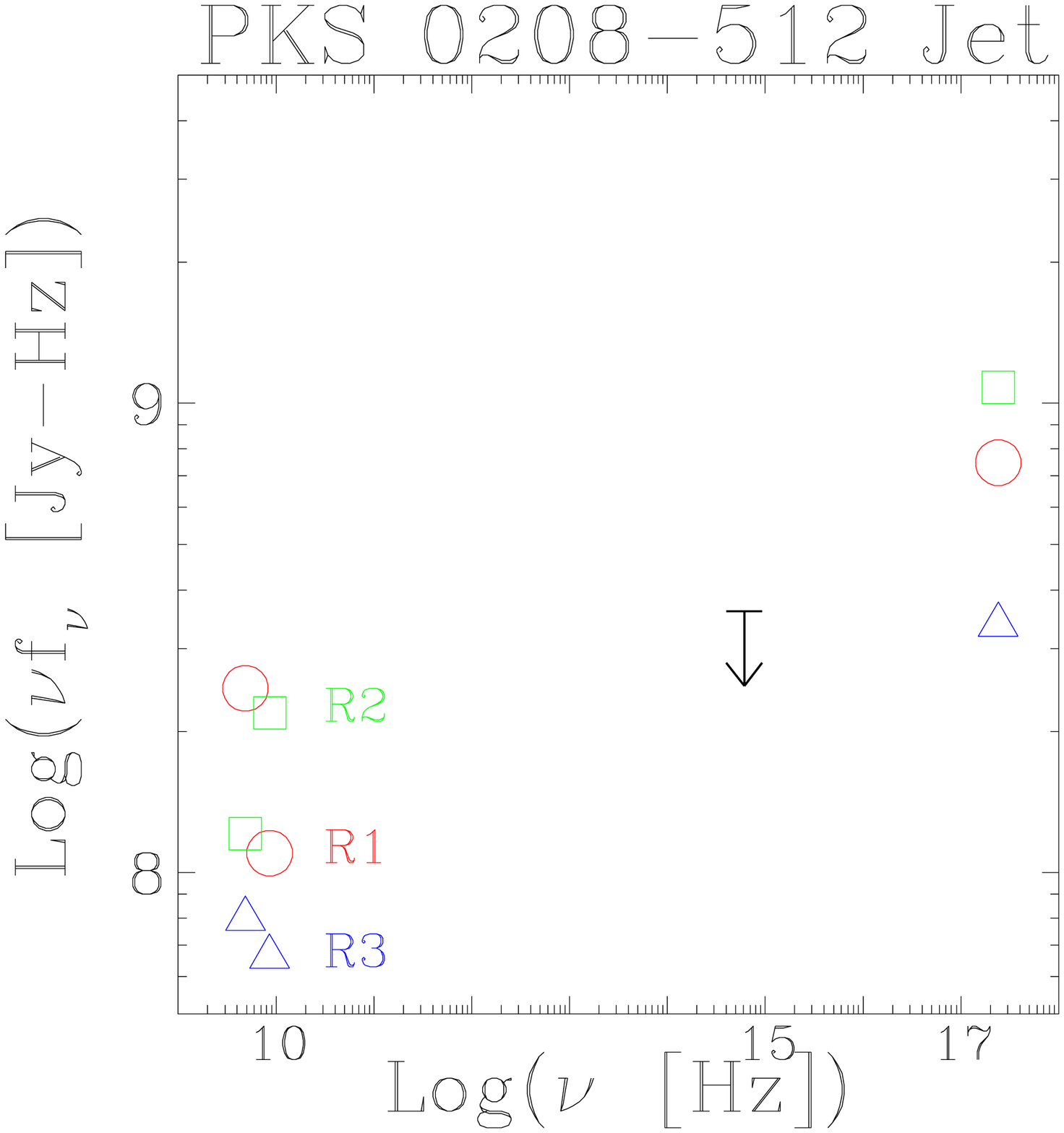}{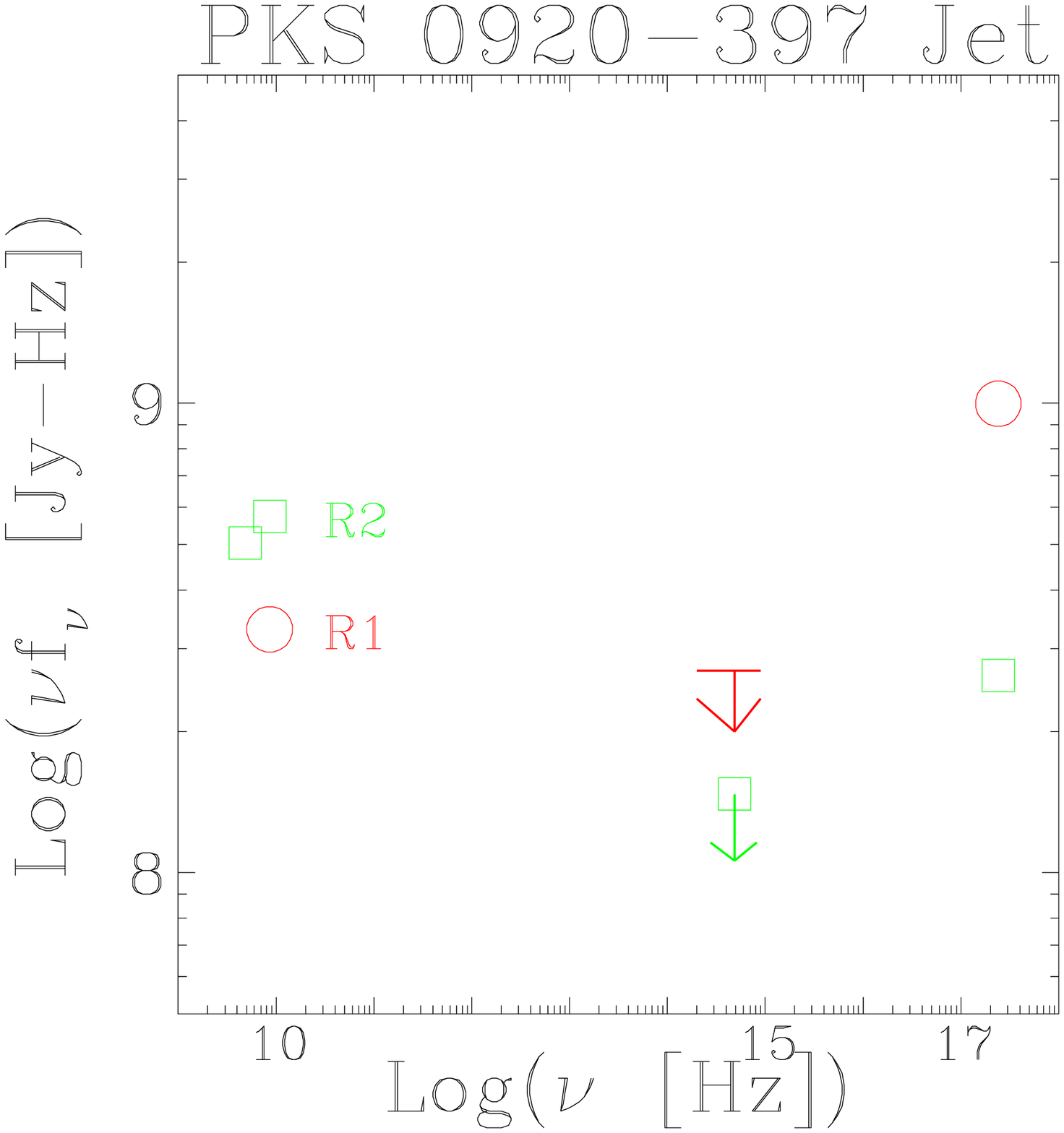} \\
\plottwo{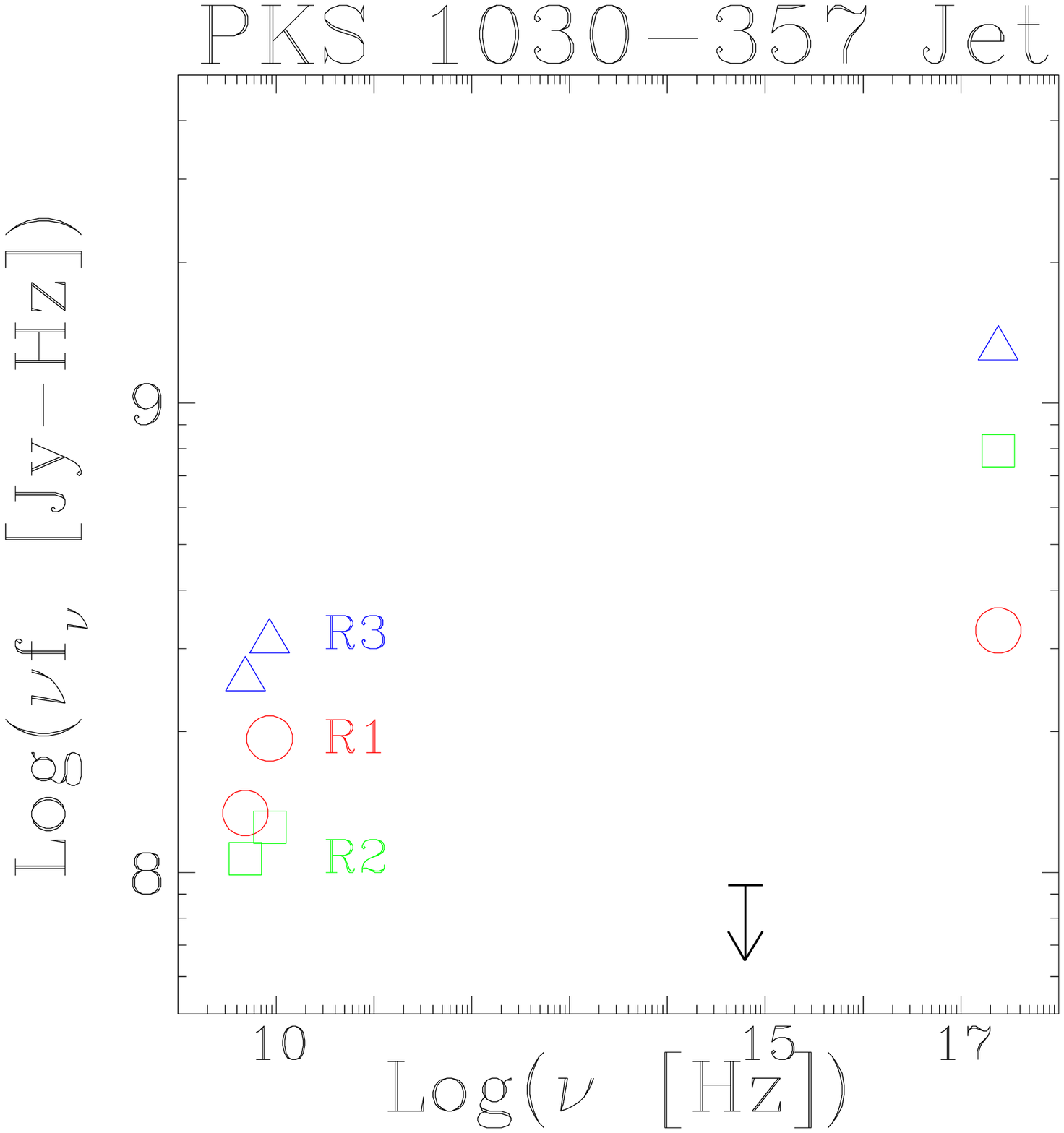}{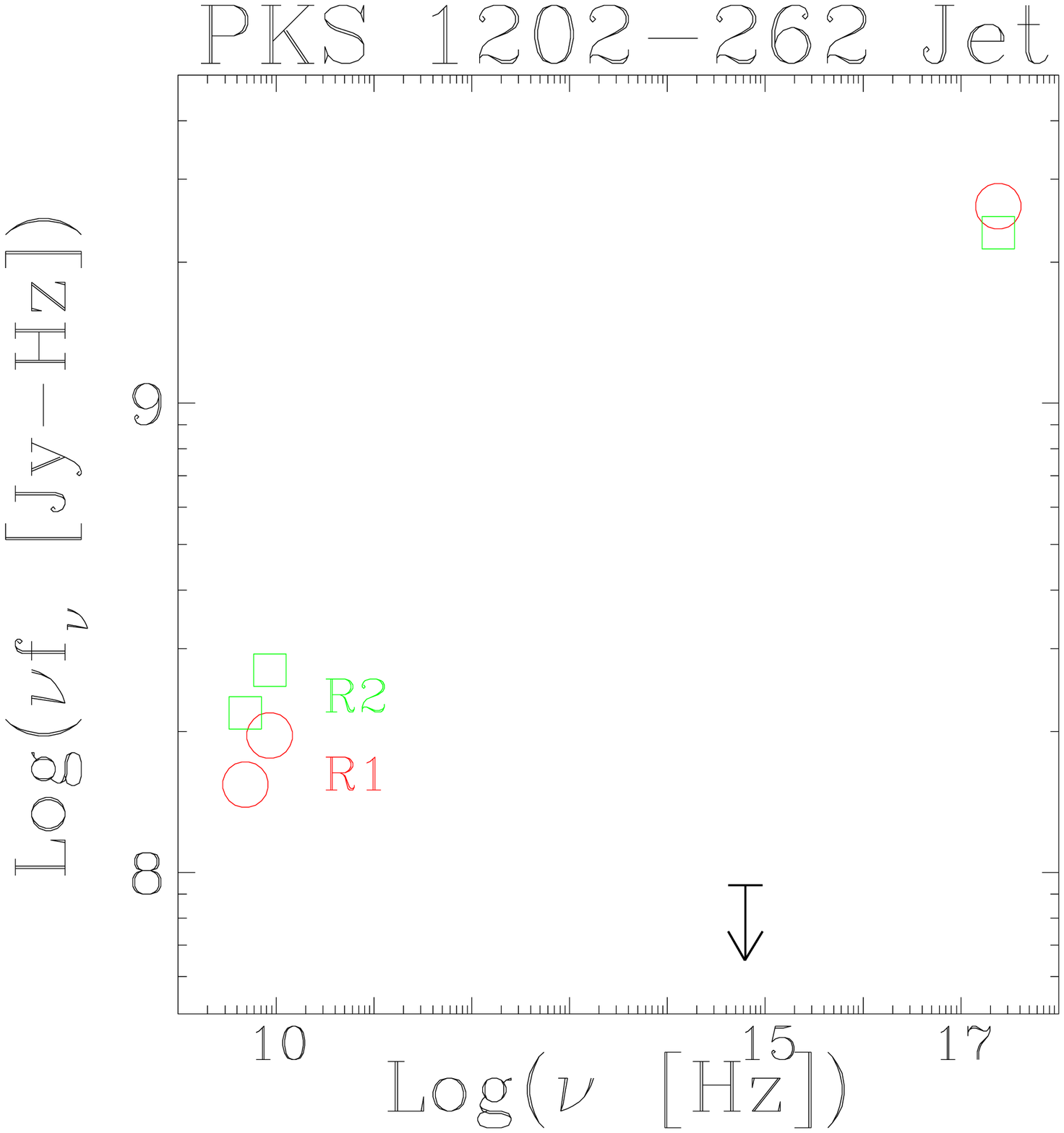}
\epsscale{1}
\caption{\label{fig:jetSED} Spectral energy distributions for the
  distinct regions of each jet. The downward arrows are our 
  upper limit to the optical flux from any region. We use a circle,
  square, and triangle, respectively, to indicate regions
  progressively further from the quasar.}
 \end{figure}

Figure~\ref{fig:jetSED} plots the spectral energy distributions for
each of the regions shown in Figure~\ref{fig:regions}. We have
obtained Magellan optical observations of the four quasars
(\citet{Gelbord03, Gelbord04}, Gelbord and Marshall, in preparation 2005).
In general no statistically significant optical emission is
detected. An exception is R2 of PKS~0920-397; however, there are other
faint optical objects in the field and we cannot rule out a chance
superposition at present. We therefore treat all optical data as upper
limits.

For PKS~0920-397, PKS~1030-357 and PKS~1202-262 our upper limit g$^{\prime}$
magnitudes  do not allow the X-rays to be a simple power-law
extrapolation of the radio synchrotron emission.  For PKS~0208-512 our
limit of 25th magnitude prevents such an extrapolation for R2. The 4.8
to 8.6 GHz radio index for PKS~0208-512 R1 and R3 also argues against
extrapolation of the synchrotron emission to the X-ray region.  

We will present the interpretation of all four objects in the context
of inverse Compton emission from the cosmic microwave background. This
has been argued to be the most plausible mechanism for the X-ray jet
in many powerful radio-loud quasars \citep{Tavecchio00, Celotti01,
Marshall01, Sambruna01,Sambruna02, Sambruna04, Harris02,
Siemiginowska02,Siemiginowska03a,Siemiginowska03b}. We discuss the
consequences and limitations of this assumption, and mention alternate
emission mechanisms in section~\ref{sec:discussion}.

\section{PHYSICAL PARAMETERS}
\label{param}

We assume that the X-ray emission from each region arises from inverse
Compton scattering by the same power-law population of electrons,
$n(\gamma)=n_0\,\gamma^{-m}$ electrons cm$^{-3}$ per unit $\gamma$,
emitting the radio-synchrotron radiation from that region.  The ratio
of synchrotron to Compton power is just the ratio of energy density of
the magnetic field to the energy density of the target photons,
assuming the latter are isotropic in the jet rest frame
\citep{Felten66}.  In applying that formalism to the powerful X-ray
jets, one typically cannot find a credible source of target photons if
one also assumes that the magnetic field and relativistic electrons
are nearly in energy equipartition, and are not in relativistic motion
\citep[e.g.,][]{Schwartz00}.  \citet{Tavecchio00} and
\citet{Celotti01} resolved this dilemma by exploiting the enhanced
apparent CMB density for electrons moving with bulk relativistic
velocity, $\beta c \approx c$, with respect to the isotropic CMB frame
\citep{Dermer94}.  In the frame of a jet moving with bulk Lorentz
factor $\Gamma$, the CMB energy density will exceed the magnetic field
energy density at redshifts
\begin{equation}
\label{eq:CMBdominates}
z\geq\max{[(0.556\sqrt{B_{\mu G}/\Gamma} -1),0]},
\end{equation}
\citep{Schwartz02}, where B$_{\mu G}$ is the magnetic field in
micro-Gauss (1$\mu$G=0.1nT), and
$\Gamma$=(1-$\beta^2$)$^{-1/2}$. Under such conditions the IC/CMB may
become the dominant energy loss mechanism for the relativistic
electrons. The energy density due to the radiation field of a quasar
emitting an isotropic, bolometric luminosity 10$^{44}\,{\rm L}_{44}$
ergs s$^{-1}$, falls below the CMB energy density at a distance 4.7
$\sqrt({\rm L}_{44}/(1+{\rm z})^2$ kpc from the quasar. For PKS
0208-512 this distance is 34 kpc, so the quasar radiation field may
play a role, especially in producing $\gamma$-ray emission. We will
consider this in more detail in connection with our deeper \axaf\ and
HST observations (Perlman et al., in preparation). For the other three
objects, the quasar field drops below the CMB energy density 7 to 14
kpc from the quasar, which is much smaller than the deprojected
distances we derive for the X-ray emitting regions.

From \citet{Felten66}, following their approximations that the flux
density at any frequency is produced by a $\delta$-function at a
characteristic mean electron $\gamma$, and that magnetic fields,
particles, and target photons are isotropic in the emitting region,
the ratio of synchrotron to 
IC/CMB emission, extrapolated to some common frequency, will be 
\begin{equation}
\label{eq:felten}
(S_{\rm synch}/S_{\rm IC}) \approx \frac{(2\times 10^4 T)^{(3-m)/2}\,
    B_{\mu G}^{(1+m)/2}}{ 8 \pi \,\rho}
\end{equation}
where $\rho= \Gamma^2 (1+z)^4 \rho_0 $ is the mean energy density
of the CMB at redshift z in a frame moving with Lorentz factor
$\Gamma$, $\rho_0$ = 4.19 $\times 10^{-13}$ ergs cm$^{-3}$ is the
local CMB energy density, and the apparent temperature of the CMB in
the jet frame is $T= (1+z) \Gamma T_0$, where the local CMB
temperature is $T_0$=2.728\arcdeg K \citep{Fixsen96}. There are two
independent parameters among the direction to our line of sight,
$\theta$, the bulk Lorentz factor, $\Gamma$, and the effective Doppler
factor $\delta$=$[\Gamma (1-\beta\cos(\theta))]^{-1}$. Since the
asymptotic value of $\delta$ is 2$\Gamma$ for large $\Gamma$ when the
jet is beamed exactly in our direction, i.e., $\cos(\theta)$=1, we
make the common assumption $\Gamma=\delta$, in the absence of any
information on $\theta$.  We assume the  spectral index
$\alpha$=(m-1)/2=0.7, and use the known CMB parameters
\citep{Fixsen96}. We know that the CMB photons are highly anisotropic
in the jet frame, but the photons we observe will be forward scattered
by electrons moving near to the line of sight, so we approximate that
they see the mean energy density. With these conditions,
equation~\ref{eq:felten} gives an estimate of the magnetic field:
\begin{equation}
\label{eq:icB}
{\rm B_{\mu G}} \approx 4.23 \times 10^{-11} \delta
 (1+z)^{2.18} (S_{\rm synch}/S_{\rm IC})^{0.588},
\end{equation}
where $S_{\rm IC}$ is the inferred X-ray flux density at 1 keV for the
assumed spectral index $\alpha$=0.7, and $S_{\rm synch}$ is the
measured radio flux density at 8.6 GHz. Appendix~\ref{app:errors}
discusses the sensitivity of our results to the spectral index. As
noted above, the regions of PKS 0208-512 are not consistent with such
a spectral index in the range 4.8 to 8.6 GHz.

Without considering relativistic beaming, we can apply the minimum
energy conditions to estimate the magnetic field \citep{Moffet75}:
\begin{equation}
\label{eq:moffet}
B_{1} \approx 328.8 (\frac{(1+k_{1})\,L_{\rm synch}\,}{\phi
  V})^{2/7}\,\, \rm{Gauss} 
\end{equation}
This formula assumes a uniform magnetic field, and isotropic particle
distribution. \citep[Other approaches to the minimum energy
formulation will result in a small dependence of the exponent on
$\alpha$; e.g.,][]{Worrall05}. We assume unit filling factor,
$\phi$=1, and a ratio of proton to electron energy density,
k$_{1}$=1. We take the lower and upper limits of the observed radio
spectrum to be $\nu_1$=10$^6$ Hz and $\nu_2$=10$^{12}$ Hz, in order to
integrate over $\nu$ to get the total synchrotron luminosity, $L_{\rm
synch}$. We consider the emitting volumes, $V$, to be cylinders with
the measured angular lengths $\theta_{l}$ and diameters
$\theta_{d}$. Appendix~\ref{app:errors} illustrates our sensitivity to
these assumptions. In the jet frame we have $B=B_{1}/\delta$
\citep{Tavecchio00, Harris02, Dermer04}, so that we have
\begin{equation}
\label{eq:synchB}
B\delta \approx  400.8 ( L_{synch} / V)^{2/7}.
\end{equation}

\begin{figure}[ht]
\plottwo{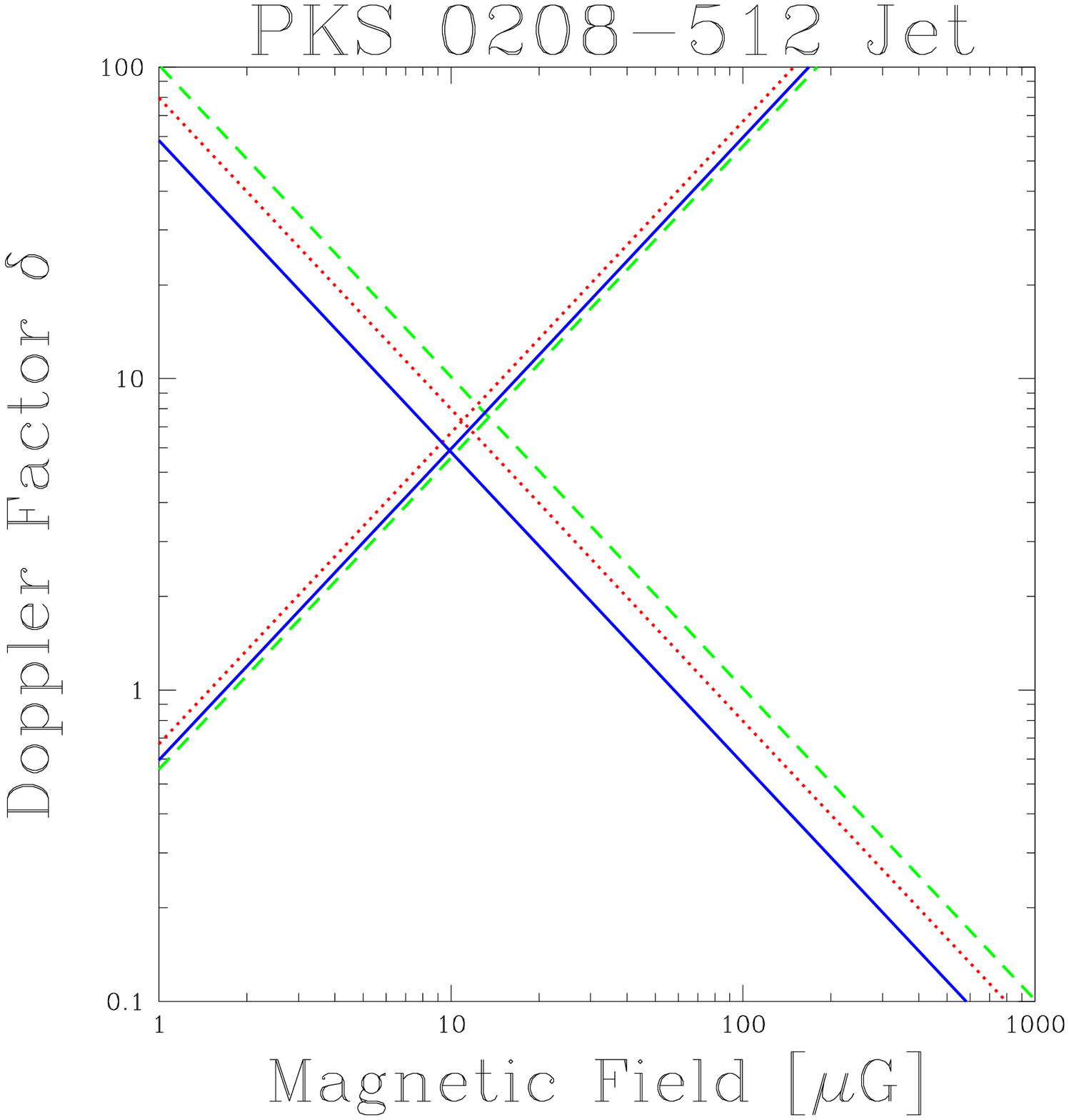}{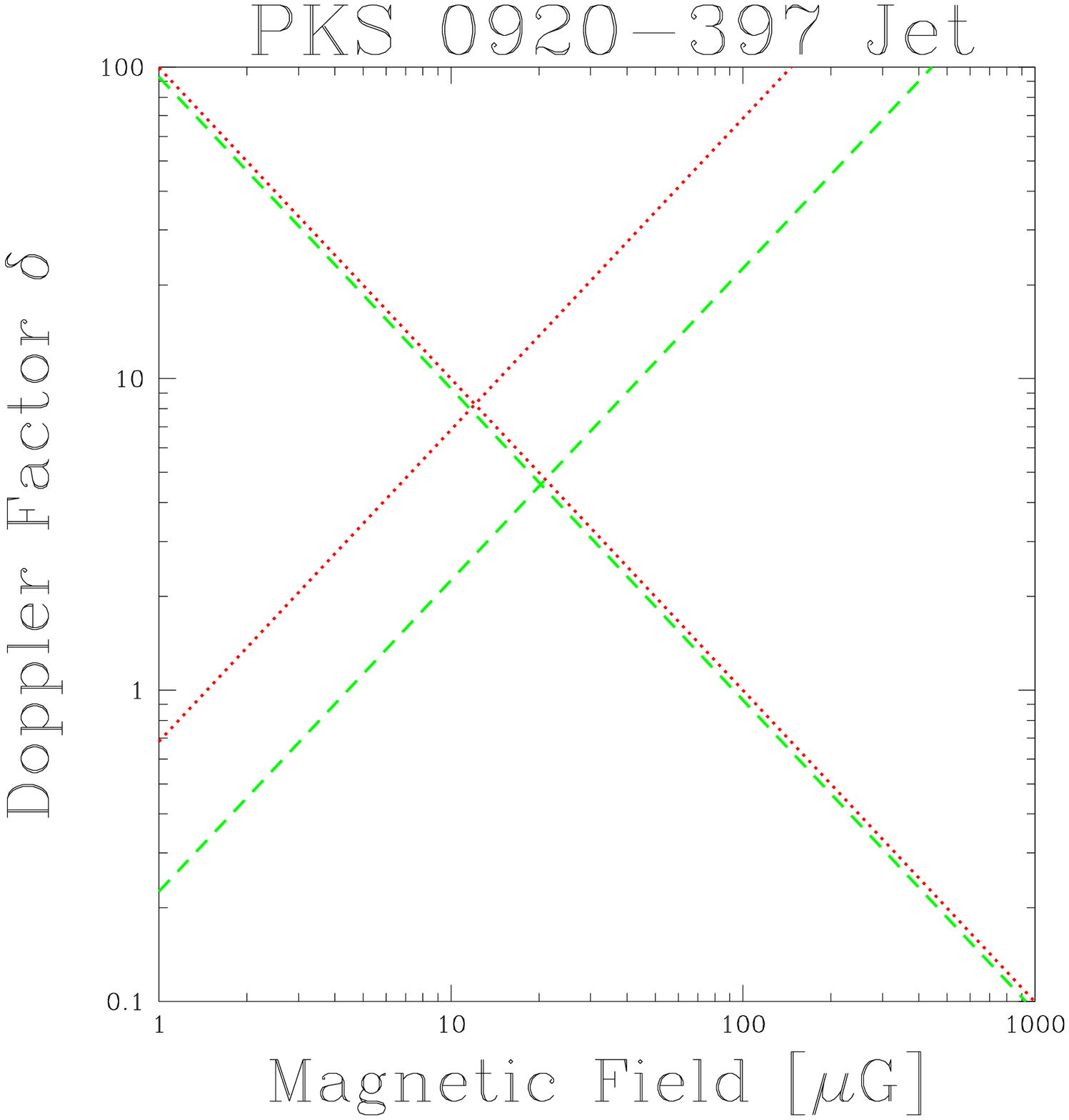}
\plottwo{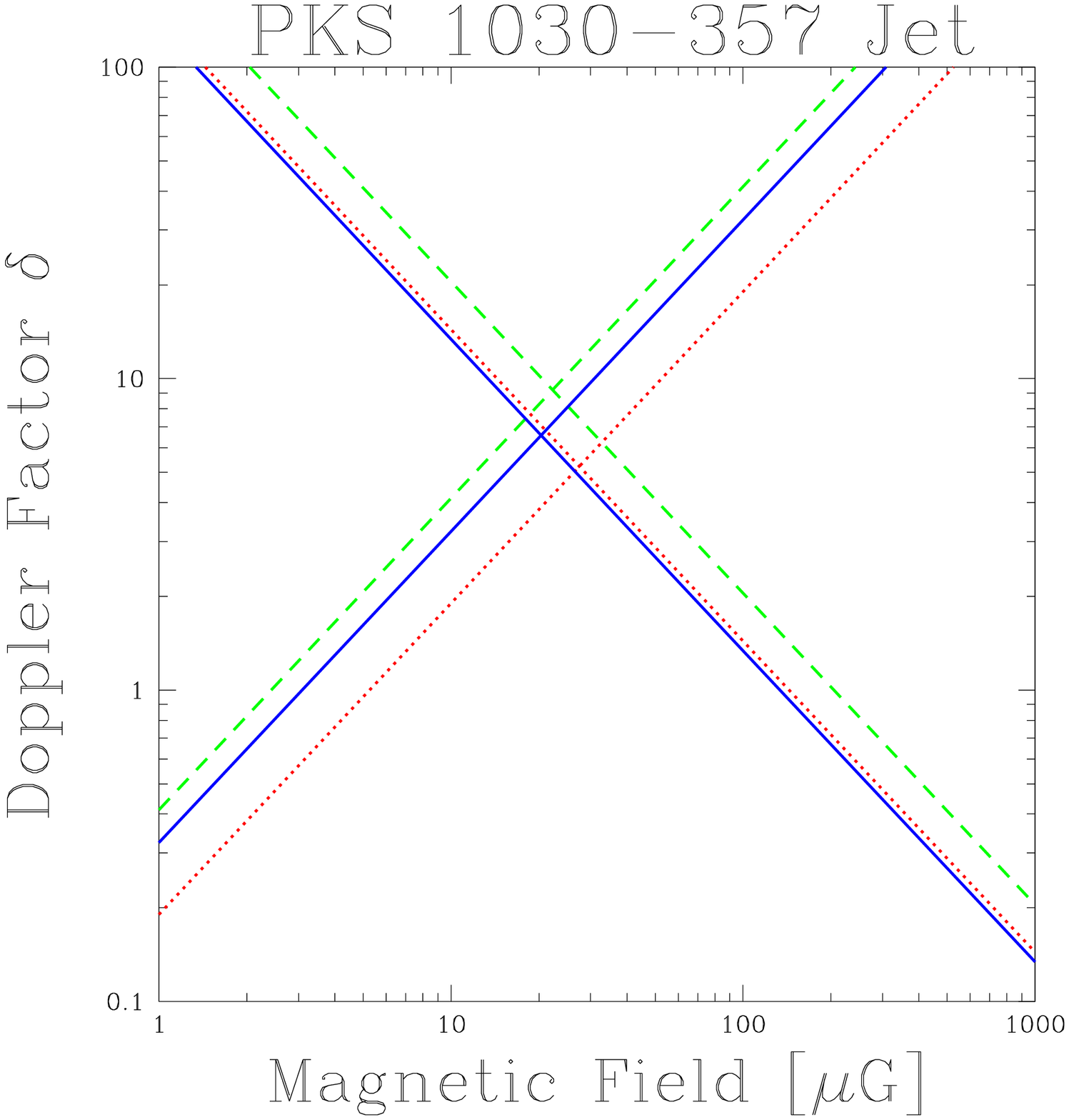}{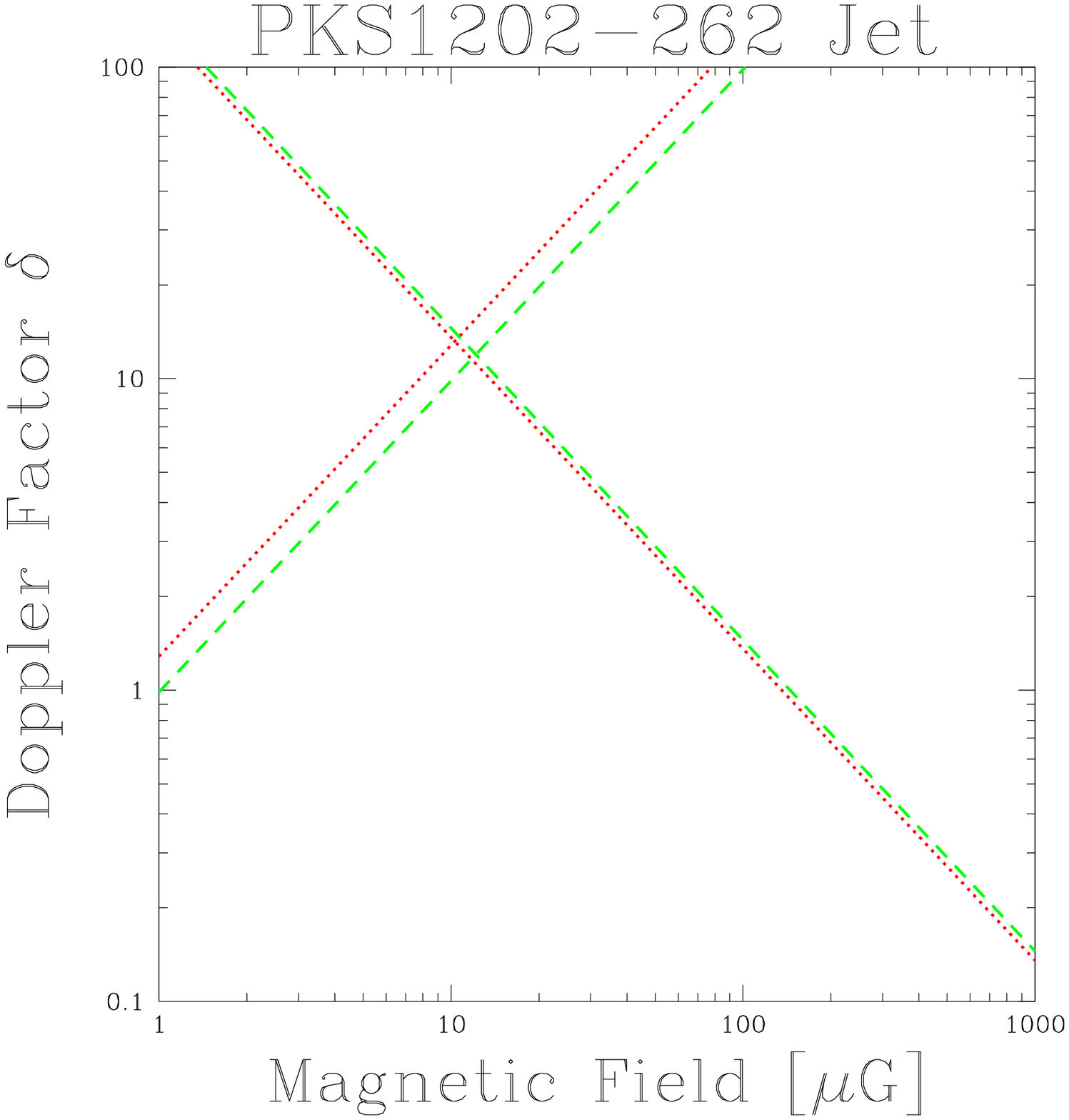}
\epsscale{1}
\caption{\label{fig:tavec} The \citet{Tavecchio00} style diagrams for the
  regions defined in Figure~\ref{fig:regions}. The intersections of
  equations \ref{eq:icB} (lines increasing with $B$)  and
  \ref{eq:synchB} (lines decreasing with $B$) give solutions for the
  Doppler factor $\delta$ and the magnetic field in the jet rest
  frame. The regions in order of increasing distance from the quasar
  are shown as dotted (``R1''), dashed (``R2''), and solid (``R3'').}
 \end{figure}  

In Figure~\ref{fig:tavec} we plot equations \ref{eq:icB} and
\ref{eq:synchB}.  The intersection of these curves for each feature in
the four jets gives a solution for the unknowns $B$ and $\delta$. These
values are tabulated in columns 3 and 4 of Table~\ref{tab:prop}. We
find magnetic fields of order 10 to 25 $\mu$G and Doppler factors in
the range 5 to 15 for the various regions in the jets. In
Appendix~\ref{app:errors} we quantify the uncertainties on those
numbers. Compared to the analysis in Paper I, which considered the
entire length of the jet and which assumed $\Gamma=10$, we here use
smaller volumes which naturally lead to larger values of the magnetic
field B$_{1}$ and smaller angles of the jet from our line of sight.

\citet{Sambruna04} have carried out a joint Chandra and HST survey,
and report the analysis of ten knots which have been detected in both
the optical and X-ray bands from six of their sources. They interpret
nine of these in terms of the IC/CMB model, and derive magnetic fields
in the range 3 to 12 $\mu$G, and Doppler factors in the range 6 to
14. The values derived here are therefore quite similar, despite some
differences in the model assumptions; e.g., the assumed shapes of the
emitting region and values of $\gamma_{min}$.  As noted by
\citet{Tavecchio04}, the values of the Doppler factor are expected to
be relatively robust since other derived quantities depend on powers
of $\delta$.

From the magnetic field strength and our arbitrary assumption of a
low-frequency cutoff at an observed $\nu$ = 10$^6$ Hz, we can
calculate the low energy electron spectrum cutoff $\gamma_{ min}$
which roughly corresponds to that frequency:
$\gamma_{min}=\sqrt((1+z)/(4.2 \delta B))$.  With that $\gamma_{min}$
we can calculate the total number density, n$_{e}$, of relativistic
electrons by equating the particle energy density with that of the
magnetic field, which is approximately the minimum energy condition:
\\ $(1+k_{1}) \int_{\gamma_{min}}^{\gamma_{max}} n_{0} m_{e} c^2
\gamma^{1-m} d\gamma \approx B^2/(8 \pi)$.  Columns 5 and 6 give
$\gamma_{min}$ and $n_{e}$, respectively.

For a fixed Doppler factor $\delta$, the maximum angle by which the
jet can deviate from our line of sight is $\arccos
[\sqrt(\delta^2-1)/\delta]$, which is also the angle for which
$\delta$=$\Gamma$ as we have assumed. From this maximum angle, given
in column 7 and the measured angular projection of the jet on the sky,
we can compute the minimum intrinsic length of each jet region, as
given in column 8 of Table~\ref{tab:prop}.

\clearpage

\begin{deluxetable}{rccccccccc}    
\footnotesize
\tablewidth{0pt}
\tablecaption{{Properties of the X-Ray Jets\label{tab:prop}} }
\tablecolumns{10}
\tablehead{
\colhead{{PKS Name}} &
\colhead{{Jet}} &
\colhead{{B}}&
 \colhead{{$\delta $}} & 
\colhead{$\gamma_{min}$\tablenotemark{b}} &
\colhead{n$_e$} &
\colhead{{$\theta_{max}$\tablenotemark{c} }} & 
\colhead{{Minimum}} &
\colhead{{Kinetic}} &
\colhead{{Radiative}} \\

\colhead{{(region)}} &
\colhead{{Frac\tablenotemark{a}  }} &
\colhead{{[$\mu$G]}}&
 \colhead{{}} & 
 \colhead{{}} &
\colhead{10$^{-8}$ cm$^{-3}$} &
\colhead{{[deg]}} & 
\colhead{{Length\tablenotemark{d}}} &
\colhead{{Flux\tablenotemark{e}  }} &
\colhead{{Efficiency\tablenotemark{f}}}
}

\startdata
{
0208-512  }&{\nodata }&{\nodata}&{
  \nodata}&{\nodata}&{\nodata}&{\nodata}&{\nodata}&{\nodata}&{\nodata}\\ 
(R1) & 0.010&10.9 &7.3 &77&  1.1& 7.8 &156 &9.5 & 0.3\\ 
(R2) & 0.015&13.5 &7.5&69&  1.8&  7.7 &262 &3.6 &1.2 \\
(R3)?  & 0.004&10.1 &5.7&91&  0.78& 10.1 &246 &3.8 & 0.5\\
 & & & & & & & \\

{0920-397}&{\nodata  }&{ \nodata}&{\nodata}&{
 \nodata}&{\nodata}&{\nodata}&{\nodata}&{\nodata}&{\nodata}\\
(R1) &0.037 &12.1 &8.3 &61& 1.7& 6.9 &322 &1.0 & 1.0\\
(R2)? &0.010 &20.8 &4.7 &62&4.8& 12.3 &356 &3.8 & 0.2\\
 & & & & & & & \\
 
{1030-357 }&{\nodata }&{ \nodata }&{ \nodata}&{\nodata }&{\nodata
}&{\nodata}&{\nodata}&{\nodata}&{\nodata} \\
(R1)? &0.018 &27.1 &5.2 & 64& 7.9& 11.1 &362 &1.7 &4.1 \\
(R2) & 0.043&22.3 &9.2 & 53& 6.5&  6.2 &1155 &3.4 & 1.8\\
(R3) &0.071 &20.9 &6.7 & 65& 4.7 & 8.6 &1484 &5.5 & 3.0\\
 & & & & & & & \\

{1202-262}&{\nodata } &{ \nodata } &{\nodata}
&{\nodata}&{\nodata}&{\nodata}&{\nodata}&{\nodata}&{\nodata}\\ 
(R1) &0.076 &10.6 &13.5 & 55& 1.4&    4.2 &443 &0.9 & 2.2\\
(R2) & 0.066&12.0 &11.8 &  55& 1.8&   4.9 &568 &4.0 & 0.6\\
 & & & & & & & \\
 
\enddata
\tablenotetext{a}{{X-ray flux in Jet divided by X-ray flux in quasar}}
\tablenotetext{b}{{Calculated from B so that electrons of
    $\gamma_{min}$ give 1 MHz synchrotron emission}}
\tablenotetext{c}{{Calculated assuming bulk Lorentz factor $\Gamma$
    equals Doppler factor $\delta$}}
\tablenotetext{d}{{Minimum distance from quasar, deprojected by 1/$\sin(\theta_{max})$, in kpc}}
\tablenotetext{e}{{Kinetic power of jet, 10$^{46}$ erg s$^{-1}$}}
\tablenotetext{f}{{De-beamed luminosity divided by kinetic flux, in 10$^{-4}$}}
\end{deluxetable}


\clearpage

The kinetic energy flux carried by the jet in our observer frame is given by 
\begin{equation}
P_{jet}= A \Gamma^2 \beta  c (w-\rho_{0} c^2/\Gamma),  
\end{equation}
 where A is the cross sectional area, $\Gamma$ is assumed equal to
$\delta$, $\rho_{0}$ is the rest mass density, and $w$ is the total
relativistic enthalpy density in the jet rest frame
\citep[e.g.,][]{Bicknell94}. Calculation of this quantity is discussed
in Appendix~\ref{app:kinetic}, and it is tabulated in column 9 of
Table~\ref{tab:prop}. It gives the ability of the jet to do work on
its surroundings, and explicitly excludes the rest mass energy of the
particles, which does not normally enter the energy budget of the
black hole and cannot normally be recovered from the jet. It differs
by a correction of order unity from the commonly used bulk power
calculated using only the internal energy density flux \citep[e.g.,][]
{Ghisellini01}.

\begin{figure}[h]
\epsscale{0.6}
\plotone{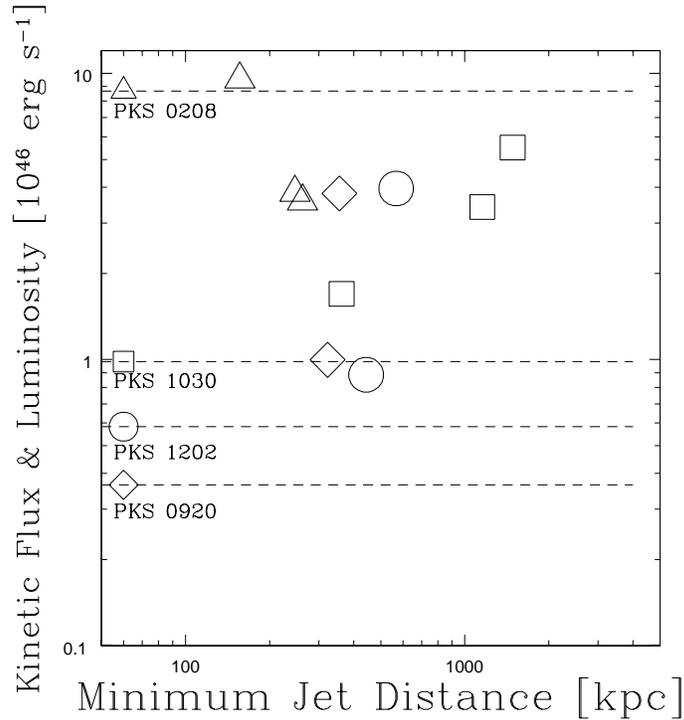}
\epsscale{1}
\caption{\label{fig:kineticflux} 
Kinetic power flowing through each
region, plotted against the minimum deprojected distance from the
quasar.  Larger symbols indicate regions of the jets in PKS 0208-512
(triangles), PKS0920-397 (diamonds), PKS 1030-357 (squares), and PKS
1202-262 (circles).  The dashed lines give the estimated radiative luminosity of
the core quasar, with the symbol and name plotted to the left.}
\end{figure}  
\clearpage

The kinetic energy flux carried by the jet in our observer frame is given by 
\begin{equation}
P_{jet}= A \Gamma^2 \beta  c (w-\rho_{0} c^2/\Gamma),  
\end{equation}
 where A is the cross sectional area, $\Gamma$ is assumed equal to
$\delta$, $\rho_{0}$ is the rest mass density, and $w$ is the total
relativistic enthalpy density in the jet rest frame
\citep[e.g.,][]{Bicknell94}. Calculation of this quantity is discussed
in Appendix~\ref{app:kinetic}, and it is tabulated in column 9 of
Table~\ref{tab:prop}. It gives the ability of the jet to do work on
its surroundings, and explicitly excludes the rest mass energy of the
particles, which does not normally enter the energy budget of the
black hole and cannot normally be recovered from the jet. It differs
by a correction of order unity from the commonly used bulk power
calculated using only the internal energy density flux \citep[e.g.,][]
{Ghisellini01}.

For a pure electron/positron jet, (i.e., $k_{1}$=0), the powers would
be a factor $\approx$ 3 lower. For the case where charge neutrality
was maintained by an equal number density of cold protons and with
$k_{1} \approx$ 10, the kinetic power would be about 20 times
larger. We compare with the quasar bolometric radiative luminosity,
estimated by 
fitting the radio loud template of \citet{Elvis94} to the optical
magnitude from the NED database, assuming isotropic emission, and integrating over all wavelengths. As shown in
Figure~\ref{fig:kineticflux}, the kinetic powers are typically of
order or larger than the bolometric accretion luminosity of the
quasar. If the core optical emission is beamed, then the intrinsic
luminosity of the cores are even smaller, and this conclusion is
strengthened. As pointed out by \citet{Meier03}, this requires that the jet
formation be considered as an essential feature of the accretion
process which is powering the quasar.  The low efficiency,
Table~\ref{tab:prop}, column 10, with which these jets radiate their
kinetic power is consistent with the ability to transport energy from
the black hole core to distant radio lobes. 

For PKS 0208-512 \citet{Maraschi03} have estimated a jet power
$\approx$ 5$\times $10$^{47}$ ergs s$^{-1}$ at a few pc from the
core. If we assume equal numbers of electrons and of cold protons, as
they do, our average kinetic power for the regions R1 and R2 would be
about 1.2$\times $10$^{48}$ ergs s$^{-1}$. This is therefore
consistent with the conclusion of \citet{Tavecchio04} who find that
the power transported to the outer jets of PKS 1510-089 and 1641+399
is similar to the energy flux traveling through the pc scale regions.

\section{DISCUSSION}
\label{sec:discussion}

The kinetic flux carried by the jet equals the core luminosity for PKS
0208-512, and dominates for the other three quasars. From the
discussion in Appendix~\ref{app:kinetic}, we have for the kinetic
energy flux  
\begin{equation}
\label{eq:kineticflux}
P_{jet} \approx 1.25 \times 10^{9}   \delta^2
B^2 (\theta_d D_A)^2 (2+\chi (\delta -1)/\delta),  
\end{equation}
where $D_A$ is the angular size distance and $\chi$ is the ratio of
rest mass energy density to particle enthalpy density, and we have
considered a tangled magnetic field with $B_{\perp}^2 = \frac{2}{3}
B^2$.  The value of $P_{jet}$ is not very sensitive to the values of
$B$ and $\delta$ as long as $\beta \approx$1, and we are near
equipartition (with $\epsilon_{\rm
electrons}$(1+k$_{1}$)=B$^2$/8$\pi$), since the product $B\delta$ is
determined by the observed and assumed quantities in
equation~\ref{eq:moffet}. The importance of the X-ray observations is
that they show the jets must be in substantial relativistic bulk
motion with $\beta \approx$1, if we explain the X-rays as due to the
IC/CMB mechanism.

Alternate models for the X-ray emission have recently been
considered. These are motivated at least in part by morphological
arguments. Because the cooling length of the IC X-ray
emitting electrons  ($\gamma \sim$  hundreds) is longer than that 
of the radio  emitting electrons ($\gamma \sim$  thousands), the
extent of the X-ray emission should always be longer than that of the
radio, in a model where the electrons are all accelerated at some
discrete locations within a jet. However, some cases clearly show an
X-ray peak upstream of a radio peak, which could be taken as evidence
that the X-rays are produced by higher energy electrons;  e.g.,
\citet{Stawarz04}. 

\citet{Dermer02} proposed that the observed X-rays in knots are due to
 synchrotron emission from electrons cooling by IC/CMB in the
 Klein--Nishina regime. This results in a high-energy ($\gamma\gtrsim
 10^8$) ``hump" in the electron distribution function, manifested as a
 hardening of the synchrotron spectrum between UV and X-ray energies
 for an appropriately low magnetic field.  In this model, the
 extrapolation of the harder X-ray spectrum to optical frequencies
 must always lie below the actual optical flux density.  This model is
 ruled out by optical detections or upper limits at several knots;
 e.g., PKS 0637--752, \citep{Chartas00}; Knot A of 1354+195,
 \citep{Sambruna04}; Knot B of 1150-089, \citep{Sambruna02}; and Knot
 C4 of 0827+243, \citep{Jorstad04}.  The limited statistics of our
 current X-ray data do not provide us with spectral information, and
 therefore we cannot check the validity of this model for the present
 quasars.

It has also been suggested \citep[e.g.,][]{Atoyan04} that the X-rays are
due to synchrotron radiation by a second, high-energy electron
population.  This requires a low-energy cut-off at a sufficiently high
energy in the second electron distribution, so that its synchrotron
emission does not over-produce the knot optical fluxes. Even if such a
cut-off can be produced, the electrons will cool to energies below it
in less than the escape time from the knot, to produce, in this
regime, an electron distribution $N_e (\gamma)\propto \gamma^{-2}$,
resulting in a $\nu^{-1/2}$ optical spectrum that in many cases, such
as in PKS 0637--752, overproduces the observed optical fluxes.
In addition, given typical X-ray spectral indices \citep[$\alpha_x \sim
0.5 - 0.8 $; e.g.,][]{Sambruna04}, and the fact that the cooling
time of the X-ray emitting electrons is faster that the knot escape
time, the injected electron distribution must have an index $p$
flatter by one unit than the steady-state knot electron distribution
index $m$, i.e., $p =m-1=2\alpha_x \sim 1-1.6 $, significantly flatter
than that predicted by particle acceleration theories \citep[$ p \simeq 2 -
2.3$, e.g.][]{Kirk00}, unless there is much \emph{in situ}
acceleration in the knot. Spatially distributed, continuous acceleration
 \citep[e.g.,][]{Jester01,Jester05,Marshall02,Stawarz04,Perlman05} may
 provide a 
solution to this problem .   Such models can
produce a pile-up of high energy electrons at the upper end of the
electron distribution which could lead to X-ray synchrotron consistent
with observations. However, similar to the model of \citet{Dermer02},
such mechanisms cannot successfully reproduce emission by sources 
in which the extrapolation of the X-ray spectrum to optical
frequencies lies above the observed optical flux.

A different type of two-component model has been proposed by
\citet{Aharonian02}. According to this model, the X-rays are the 
synchrotron radiation of ultra high energy protons ($\gamma_p\gtrsim
10^9$). This requires a magnetic field $B \gtrsim 1 \; \rm{mG}$,
larger by a factor of $\sim 50$ than those typically used in leptonic
models, and that the propagation of protons in the knot environment is
taking place close to the Bohm diffusion limit. According to this
model, the radio-optical component is synchrotron emission from an
electron population injected in the knot, at a level which must be carefully
fine-tuned to reproduce the radio-optical continuum.

Although our objections to the above alternate explanations may not be
decisive for these four quasars, the IC/CMB approach should be
considered the simplest explanation as it adds only one parameter to
the common assumption of minimum energy. The morphological data are
certainly relevant and must be explained; however, they should not at
present be considered as contradictory to the IC/CMB approach for two
reasons. First, for the case that a radio peak appears downstream of
an X-ray peak, we know that the naive model in which electrons are
accelerated at a discrete location and then are advected downstream
can not be directly applied, since in that case the peak emission must
be coincident in both bands \citep{Hardcastle03}, or at most be offset
by an amount small compared to the angular resolution. Second, the cases
where the X-ray emission peaks closer to the core, gradually
decreasing outward, while the radio emission increases outward to peak
practically at the jet terminus, might be explained if the large-scale
jet gradually decelerates \citep{Georganopoulos04} downstream from the
first knot. The X-ray brightness then decreases along the jet because
the CMB photon energy density in the flow frame decreases.  At the
same time, the deceleration leads to an increase of the magnetic field
in the flow frame, which enhances the radio emission with distance. As
a result the radio emission is shifted downstream of the X-rays and
$\alpha_{rx}$ increases along the jet. As shown in Figure 1g of Paper
I, PKS~0920--397 is similar to 3C~273 \citep{Marshall01} in displaying
such morphology and we note that our modeling (Table~\ref{tab:prop})
of PKS 0920--397 gives a decreasing Doppler factor and increasing
magnetic field going from R1 to R2, (but see Appendix~\ref{app:errors}
for systematic uncertainties).

If the IC/CMB mechanism proves relevant to jet X-ray production, it
may have the important cosmological implication that jets with
identical intrinsic parameters will appear to have the same surface
brightness at whatever redshift they might exist
\citep{Schwartz02}. We note for the present jets, that even if they
are not in relativistic motion and that their X-ray emission is not
IC/CMB, in the equipartition models we have assumed the magnetic
fields are in the range 50 
to 100 $\mu$G, (the product B$\delta$ from columns 3 and 4 of
Table~\ref{tab:prop}). For such magnetic fields, at redshifts larger
than 3 to 4.5, respectively,  the
CMB will dominate the target photon energy density, and the
predominant X-ray mechanism \emph{will} be IC/CMB, providing the
relativistic electron spectrum extends to sufficiently low Lorentz
factors.  The present jets
would only be about 15 to 75 times fainter at such redshifts, so most
would still be detectable in long but feasible \axaf\ observations of
a few 100 ks.

Improved radio imaging and spectra, as well as deeper X-ray
observations, will allow a more quantitative confrontation of the
alternate X-ray emission mechanisms. If we hypothesize that these jets
are carrying a constant kinetic flux we expect the (B,$\delta$) to lie
along a line of constant B$\delta$ if the average injection energy has
been relatively constant. With radio spectral indices measured to
$\pm$0.05 we could distinguish $\delta$'s differing by about 2 (see
Figure~\ref{fig:errors}). With similar constraints on the X-ray
spectral indices, we could test the hypothesis that radio and X-ray
arise from the same population of electrons. We hope to obtain such
data in our upcoming ATCA and \axaf\ observations.

\acknowledgments This work was supported by NASA contract NAS8-39073
to the \emph{Chandra} X-ray Center, and SAO SV1-61010 to MIT, and NASA
grant GO2-3151C to SAO. E.S.P. acknowledges support from NASA LTSA
grant NAG5-9997, and \axaf\ grant G02-3151B. E.S.P. and M.G. also
received support from HST and \axaf\ grants STGO-10002.01 and
G04-5107X.  Part of this research was performed at the Jet Propulsion
Laboratory, California Institute of Technology, under contract to
NASA.  We thank Dan Harris and Laura Maraschi for discussions and
comments. This research has made use of NASA's Astrophysics Data
System Bibliographic Services, and the NASA/IPAC Extragalactic
Database (NED) which is operated by the Jet Propulsion Laboratory,
California Institute of Technology, under contract with the National
Aeronautics and Space Administration.

\appendix

\section{SYSTEMATIC UNCERTAINTIES IN DETERMINING B AND $\delta$}
\label{app:errors}

\begin{figure}[h]
\epsscale{0.6}
\plotone{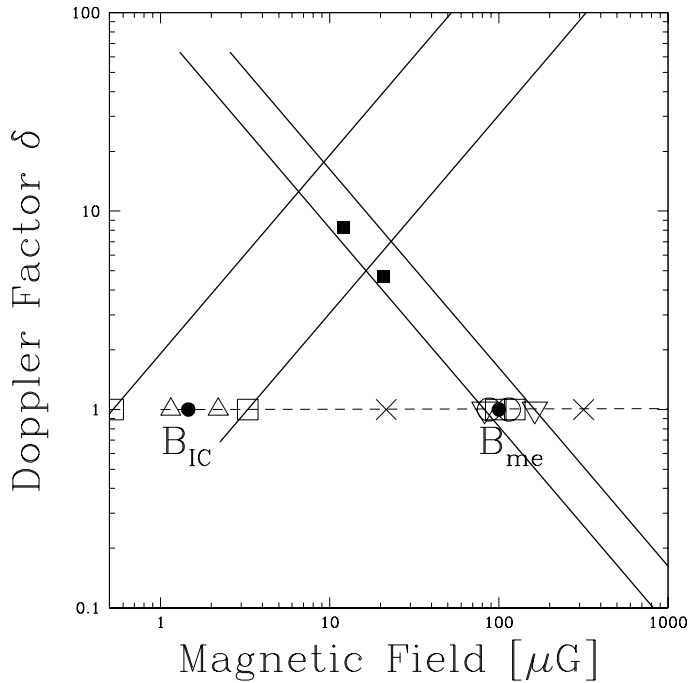}
\epsscale{1}
\caption{\label{fig:errors} 
An uncertainty analysis applied to the region R1 of PKS 0920-397. Solid dots 
give the magnetic fields according to equations \ref{eq:icB} and
\ref{eq:synchB} and for $\delta$=1. Alternate values of the
minimum energy magnetic field, $B_{me}$, or the field to produce
X-rays via IC/CMB, $B_{IC}$,
arise as follows: Crosses, internal energy 10 times larger than minimum;
open circles, low frequency radio cutoff $\nu_{1}$ = 10$^5$ or 10$^7$;
downtriangles, ratio of proton energy to electron energy $k_{1}$= 0 or 10;
open squares, electron energy index = 2.2 or 2.8; uptriangles, X-ray
flux error of a factor of 2. The filled squares are the solutions
given in Table \ref{tab:prop} for regions R1 ($B\approx$ 12) and
R2~($B\approx$ 20). The enclosed rectangle about the point $B=12.1\mu$G and
$\delta$=8.3 gives our adopted  region for the systematic uncertainty.}
\end{figure}

We consider the uncertainty in the values we derive for $B$ and
$\delta$.  We will apply equations \ref{eq:icB} and \ref{eq:synchB} to
calculate alternate values of magnetic field for the case
$\delta$=1. We vary one parameter at a time to evaluate the range
which each equation may give, and then transform the extremes by 
$B\propto\,\delta$ or $B \propto\, 1/\delta$, respectively, to give a
rectangular region representing the systematic uncertainty. This is
intended to give a conservative error; however, since we cannot
quantify the probability of the assumptions, we cannot assign a
numerical confidence.  We will illustrate by considering if we can
distinguish the Doppler factors for the distinct regions R1 and R2 in
PKS 0920-397. These are plotted as the solid squares in
Figure~\ref{fig:errors}.  We will refer to this figure in the
following discussion of systematic uncertainties.

The largest uncertainly arises from the starting assumption of
minimum energy. For lower magnetic fields the energy density U
increases $\propto \, B^{-3/2}$ and for larger fields increases
$\propto \, B^{2}$. For energy 10 times the minimum, the crosses give
the resulting fields. If conditions are far from equipartition, the
energy may be arbitrarily larger and we would have no constraints on
the magnetic field. In such circumstances it is possible that the X-rays
are produced by a different model; hence, we will not use this limit
in constructing an uncertainly region for our assumed model conditions.

In applying equation~\ref{eq:moffet} we have used assumed values of the 
radio spectral index, the lower and upper radio spectrum cutoffs, the
ratio of proton to electron energy densities, and the volume filling
factor. The squares show the effect of varying the spectral
index $\alpha$ from 0.6 to 0.9. The open circles are for low frequency
cutoffs of $\nu_{1}$ = 10$^5$ and 10$^7$. Varying the upper frequency
cutoff has an extremely small effect because the bulk of the energy is 
in the lowest energy electrons. The largest effect is due to varying
the ratio of proton to electron energy density, which we assumed as
equal, to the values $k_{1}$= 0 and 10, shown by the downward
triangles. Similar errors would arise from a 50\% error in the
cylinder radius (which in turn is merely an assumed geometry). These
define the extremes we will take for the uncertainty 
on the magnetic field estimated via the equipartition assumption. If
the filling factor were an order of magnitude smaller, $\phi
\approx$0.1, $B$ would be about twice as large, similar to the
$k_{1}$=10 case.   

The uncertainty in the magnetic field estimated from the inverse
Compton formalism is dominated by uncertainty in the radio spectral
index. These are shown by the open squares at $B\approx$0.5 and 3.3
for $\alpha$ = 0.5 and 0.9, respectively.  The upward triangles show the
effect of a factor of 2 uncertainty in the X-ray flux, due to the
Poisson errors and the uncertainty of converting counts to X-ray flux
density when we cannot measure the spectrum accurately.  
\clearpage

For PKS 0920-397 we see that the net uncertainty in
($B,\,\delta$) space about the solution for R1 almost extends to
the R2 solution. Obviously the joint uncertainty regions would 
overlap. If we assume an independent structure for these two regions,
then a single value of $B$ and $\delta$ could be assigned. On the
contrary, if the electron spectra are assumed to have the same shape
and no other parameters varied greatly between the regions,
then the uncertainties would be correlated and we would have evidence 
for deceleration of the jet.

\section{CALCULATION OF THE KINETIC ENERGY FLUX}

\label{app:kinetic}
There are various expressions in the literature for the Poynting flux
and particle kinetic luminosities of relativistic jets. Some are
incomplete and some are incorrect. One common misconception, for
example, is that the energy flux density associated with the magnetic
fields is equal to the Lorentz factor squared times the magnetic
energy density times the velocity.  Here we present the correct
equations for jet power. Although these expressions result in factor
of order unity corrections to commonly used expressions, it is useful
to have precise expressions for these quantities to provide a common
basis for comparisons of expressions for jet power.

Following \citet{Bicknell94}, let $e$ be the total (rest-mass plus
internal) energy density, $\rho_0$ the rest-mass density, $w=e+p$ the
relativistic enthalpy density, $\beta = v/c$ the bulk jet velocity and
$\Gamma = (1-\beta^2)^{-1/2}$ the bulk Lorentz factor. Then, in the
most general case, the jet energy flux density, in the absence of
magnetic fields, is given by
\begin{equation}
F_{E,i} = \left[ w \Gamma^2  - \Gamma \rho_0 c^2 \right] \, c \beta_i 
\label{e:fe_bzero}
\end{equation}
This first term is derived from the stress energy tensor of a
relativistic ideal fluid. The second term subtracts the flux of
rest-mass energy, which for the purposes of energy conversion in the
external regions of AGN, is not generally relevant.\footnote{In a pure
electron/positron jet the rest mass might be recovered via
annihilation, and the last term in equation~\ref{e:fe_bzero} should be
dropped} Note however, that the rest--mass energy contributes to the
total energy flux, through its appearance in the the relativistic
enthalpy, $w$.

We consider the jet plasma to be a
mixture of relativistically cold matter ($p << e$) and relativistic
particles with 
pressure $p = \frac{1}{3} \epsilon$, where $\epsilon = e - \rho_0 c^2$
is the internal energy density. The energy flux density becomes: 
\begin{equation} 
F_{E,i} = \Gamma^2 c \beta_i \left[ \frac{4}{3} \epsilon  + \frac{\Gamma - 1}{\Gamma} \rho_0 c^2 \right]  
\end{equation}

The synchrotron and inverse Compton radiation emitted by a jet depends
upon the relativistic electron and positron energy density, so that we
express the internal energy in terms of these quantities, and write:
\begin{eqnarray}
\epsilon &=&(1+ k_1)  \, \epsilon_e  \\
\rho_0 c^2 &=& (1+k_2)   \,  n_e m_e c^2
\end{eqnarray}
where $\epsilon_e$, $n_e$ and $m_e$ are the electron/positron energy
density, number density and mass respectively and the parameters $k_1$
and $k_2$ represent the contributions from ions to the internal energy
density and rest--mass density. For an electron--positron jet in which
only relativistic particles are present, $k_1=k_2=0$.  For an
electron--ion jet, $k_1$, in principle could take a number of values,
depending on the details of acceleration and re-acceleration and
\begin{equation}
1+k_2 = \mu \left[ 1+ \frac {\Sigma n_i  m_i}{n_e m_e} \right] \,
\frac {u}{m_e}  
\end{equation} 
where $\mu$ is the mean molecular weight, $n_i$ is the ionic number
density for an ion of mass $m_i$ and $u$ is an atomic mass unit. For a
plasma with normal cosmic abundances, $1+k_2 \approx 2,200$.

In terms of this parameterization, the energy flux density may be
represented as
\begin{equation}
F_{E,i} = \Gamma^2 c \beta_i 
\left[ \frac {4}{3} (1+k_1) \epsilon_e + \frac {\Gamma-1}{\Gamma} (1+k_2) 
n_e m_e c^2  \right]
\end{equation}
and, introducing the jet area A, the jet power (energy flux) attributed to the rest--energy density and internal energy density of the particles, is given by
\begin{equation}
P_{\rm jet}^{\rm p} \approx \Gamma^2 c \beta A \,
\left[ \frac {4}{3} (1+k_1) \epsilon_e + \frac {\Gamma-1}{\Gamma} (1+k_2) 
n_e m_e c^2  \right]
\end{equation}

The rest--mass energy density of the relativistic electrons ($n_e m_e
c^2$) may be determined from their energy density as follows. Suppose
that the number density per unit Lorentz factor of the relativistic
electrons are represented by a power-law in Lorentz factor, $\gamma$,
for $\gamma_{\rm min} < \gamma < \gamma_{\rm max}$, by
\begin{equation}
n( \gamma) = n_0 \gamma^{-a}
\end{equation}
with $a>2$. Then,
\begin{eqnarray}
n_e &=& \frac{  n_0 \gamma_{\rm min}^{-(a-1)} } {a-1} 
\left[1 - \left( \frac {\gamma_{\rm max}}{\gamma_{\rm min}}
  \right)^{-(a-1)} \right] \\ 
\epsilon_e &=& \frac { n_0 m_e c^2 \gamma_{\rm min}^{-(a-2)}}{a-2} \,
\left [ 1- \left( \frac {\gamma_{\rm max}}{\gamma_{\rm min}}
  \right)^{-(a-2)} \right] \\ 
\frac {n_e m_e c^2}{\epsilon_e} &=& \left( \frac {a-2}{a-1} \right) \,
\gamma_{\rm min}^{-1} 
\, \left[ \frac {1- (\gamma_{\rm max}/\gamma_{\rm min})^{-(a-1)}}
{1 - (\gamma_{\rm max}/\gamma_{\rm min})^{-(a-2)}} \right] 
\end{eqnarray}
For $\gamma_{\rm max} \gg \gamma_{\rm min}$, $n_e m_e c^2 / \epsilon_e \approx
(a-2)(a-1)^{-1} \gamma_{\rm min}^{-1}$. 

Note that the rest mass energy density of relativistic electrons and
the internal electron energy density are only of the same order when
$\gamma_{\rm min} \sim 1$. However, for a jet of normal cosmic
composition, the rest--mass contribution to the jet power is
comparable to or greater than the internal energy contribution for
$\gamma_{\rm min} \lessapprox 1000$.

To calculate the Poynting flux we use a Cartesian coordinate system defined
such that the $x$--axis is in the direction of the jet flow, the
$y$--axis is perpendicular to the flow and the magnetic field is in
the $x-y$ plane.  

Let \textbf{E} and \textbf{B} be the electric and
magnetic field in the lab--frame, with the corresponding fields
\begin{math} \mathbf{E'} \end{math} and \begin{math} \mathbf{B'} =
  B_\parallel^\prime \mathbf{\hat x} + B_\perp^\prime \mathbf{\hat y} 
\end{math} in the jet rest--frame. As a result of the high
conductivity of the plasma, $\mathbf{E^\prime = 0}$.  
Therefore, in this coordinate system, the transformation from rest
frame to lab frame (see  \citet{jackson75}) is given by 

\begin{eqnarray}
\mathbf{E} & = & \Gamma \beta B^\prime_{\perp} \mathbf{\hat{z}}\\
\hbox{and} \hspace{0.6cm} \mathbf{B} & = & B^\prime_{||}
\mathbf{\hat{x}} - \Gamma B^\prime_{\perp} \mathbf{\hat{y}} 
\end{eqnarray}

The component of Poynting flux along the jet in cgs units is
\begin{eqnarray}
S_{x} & = & \frac{c}{4 \pi} \left( \mathbf{E} \times \mathbf{B}
\right) \cdot \mathbf{\hat{x}} 
\nonumber \\
& = & \Gamma^2 \beta c \left( \frac{B_{\perp}^{\prime 2}}{4 \pi} \right)
\end{eqnarray}

For the case of a tangled magnetic field with uniform strength, then
the average squared perpendicular magnetic field $\langle B_\perp^2
\rangle = 2/3 \langle B^2 \rangle$ and the area---integrated jet
electromagnetic power is given by:
\begin{equation}
P^{EM}_{\rm jet} \approx A \Gamma^2 \beta c \left( \frac{B^{\prime
    2}}{6 \pi} \right)  
\end{equation}
However, conservation of magnetic flux in a smooth expanding jet flow
implies that $\mathbf{B}^\prime_{\perp}$ will dominate over
$\mathbf{B}^\prime_{||}$ in the outer jet. In that case we expect
\begin{equation}
P_{jet}^{\rm EM} \approx  A \Gamma^2 \beta c 
\left( \frac{B^{\prime 2}}{4 \pi} \right) 
\end{equation}

In view of the above, the total jet power is given by:
\begin{equation}
\label{eq:totalpower}
P_{\rm jet} \approx 
\Gamma^2 c \beta A 
\left[
 \frac {4}{3} (1+k_1) \epsilon_e + 
 \frac {\langle B_\perp^{\prime 2} \rangle }{4 \pi} +
 \frac {\Gamma-1}{\Gamma} (1+k_2) n_e m_e c^2 
\right]
\end{equation}
If the internal energy density and magnetic field are in equipartition, then
$\langle B^{\prime 2} \rangle /8 \pi \approx (1+k_1)\epsilon_e  $.

\end{document}